\documentclass[prl,twocolumn,amsmath,amssymb,showpacs,superscriptaddress]{revtex4-2}
\usepackage{graphicx}
\usepackage{epstopdf}
\usepackage{dcolumn}
\usepackage{bm}
\usepackage[FIGTOPCAP]{subfigure}
\usepackage[usenames]{color}
\usepackage{mathtools}
\usepackage[all]{xy}
\def \bea{\begin{eqnarray}}
\def \eea{\end{eqnarray}}

\begin{document}
\title{Disorder-Induced Long-Ranged Correlations in Scalar Active Matter}

\author{Sunghan Ro}
\affiliation{Department of Physics,
Technion-Israel Institute of Technology,
Haifa, 3200003, Israel.}
\author{Yariv Kafri}
\affiliation{Department of Physics,
Technion-Israel Institute of Technology,
Haifa, 3200003, Israel.}
\author{Mehran Kardar}
\affiliation{Department of Physics,
Massachusetts Institute of Technology,
Cambridge, Massachusetts 02139, USA.}
\author{Julien Tailleur}
\affiliation{Universit{\' e} de Paris, laboratoire Matière et Systèmes Complexes (MSC),
UMR 7057 CNRS, 75205 Paris, France.
}

\begin{abstract}
We study the impact of quenched random potentials and torques on scalar active matter. Microscopic simulations reveal that motility-induced phase separation is replaced in two-dimensions by an asymptotically homogeneous phase with anomalous long-ranged correlations and non-vanishing steady-state currents. Using a combination of phenomenological models and a field-theoretical treatment, we show the existence of a lower-critical dimension, $d_c=4$, below which phase separation is only observed for systems smaller than an Imry-Ma length-scale. We identify a weak-disorder regime in which the structure factor scales as $S(q) \sim 1/q^2$ which accounts for our numerics. In $d=2$ we predict that, at larger scales, the behaviour should cross over to a strong-disorder regime. In $d>2$, these two regimes exist separately, depending on the strength of the potential.
\end{abstract}


\maketitle
The influence of disorder on active systems has attracted a lot of interest recently~{\cite{chepizhko_diffusion_2013,reichhardt_active_2014,bechinger_active_2016,pince_disorder-mediated_2016,morin_diffusion_2017,sandor_dynamic_2017,peruani_cold_2018,reichhardt_clogging_2018,stoop_clogging_2018,ben_dor_ramifications_2019,doussal2020velocity,bhattacharjee_confinement_2019,bhattacharjee_bacterial_2019-1,woillez2020active}}. In particular, long-range order has shown a surprising stability against the introduction of quenched disorder~\cite{das_polar_2018,toner_hydrodynamic_2018,
toner_swarming_2018,maitra_active_2020,chepizhko_optimal_2013,morin_distortion_2017,chardac_meandering_2020}.
For systems belonging to the Vicsek universality class, where the order parameter has a continuous symmetry, the lower critical dimension was shown to be $d_c=2$: long-ranged polar order is observed in $d = 3$ and quasi-long-ranged order in $d = 2$~\cite{toner_hydrodynamic_2018,toner_swarming_2018,maitra_active_2020}.
This makes such active systems {\it more robust} to disorder than equilibrium ones with a continuous symmetry, for which $d_c = 4$~\cite{imry_random-field_1975,aharony_lowering_1976,fisher_ising_1984,belanger_random-field_1983,imbrie_lower_1984,bricmont_lower_1987,glaus_correlations_1986}.  

While a lot of effort has been devoted to polar aligning active matter, comparatively less is known on the influence of disorder on the collective properties of scalar active matter, when the sole hydrodynamic mode is the density field. There,  the combination of self-propulsion and kinetic hindrance leads to motility-induced phase separation (MIPS), even in the absence of attractive interactions, in dimensions $d \geq 2$~\cite{tailleur2008statistical,fily_athermal_2012,buttinoni_dynamical_2013,cates_when_2013,stenhammar_continuum_2013,redner_structure_2013,solon_active_2015,redner_classical_2016,paliwal_chemical_2018,cates_motility-induced_2015,solon_generalized_2018,solon_generalized_2018-1,tjhung_cluster_2018,thompson_lattice_2011,kourbane-houssene_exact_2018,whitelam_phase_2018,palacci_living_2013,geyer_freezing_2019}. Despite important differences, MIPS shares many features of an equilibrium liquid-gas phase separation. The latter is  stable to disorder above a lower-critical dimension $d_c=2$, and it is natural to ask whether the same holds for MIPS. 

\begin{figure} [b]
	\center
	\includegraphics[width=1.0\linewidth]{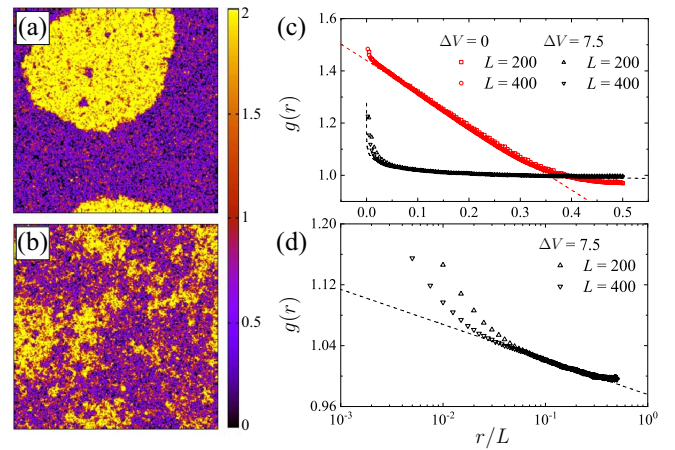}
	\caption{Snapshots of simulations without {\bf (a)} and with {\bf (b)} disorder. Color encodes density, obtained by averaging occupancies over $4$ neighbouring sites. {\bf (c)} The pair correlation functions are shown in linear scale with (black) and without (red) disorder. {\bf (d)} Pair correlation with disorder using log-linear scale. The dashed lines correspond to linear (red) and logarithmic (black) decays. Parameters: $L=300$ (a-b), $\Delta V=7.5$ (b), $v=13$, $\alpha = 1$, $n_M = 2$, $\rho_0 \equiv N/L^2=1$.}
	\label{Fig1}
\end{figure}  

In this Letter, we address this question by studying model systems of
scalar active matter in the presence of quenched random potentials and
torques using a combination of analytical and numerical
approaches. The relevance of our results to experimental systems is
discussed in the conclusion.  We show that MIPS is destroyed for
$d \leq d_c$ with $d_c=4$: The system only looks phase separated below
an Imry-Ma length scale. Instead, disorder leads to asymptotically
homogeneous systems with persistent steady-state currents. For $d>2$,
the system is either in a weak-disorder regime or in a strong-disorder
one depending on the strength of the random potential. In the
weak-disorder regime, the system is shown to exhibit self-similar
correlations with a structure factor decaying as a power law,
$S(q) \sim q^{-2}$, at small wave numbers $q$.  This behavior is very
different from that of an equilibrium scalar system, where
correlations are short-ranged with a structure factor behaving as a
Lorentzian squared~\cite{glaus_correlations_1986,kardar_2007}.  In
$d=2$, we instead predict a crossover between weak- and
strong-disorder regimes at a length scale that we
identify. Numerically we only observe the weak-disorder regime, in
which we measure a pair-correlation function that decays
logarithmically, in agreement with our analytical
predictions. Interestingly our results show that, contrary to what was
reported for the transition to collective
motion~\cite{toner_hydrodynamic_2018,toner_swarming_2018}, scalar
active systems are more fragile to disorder than passive ones. Our
results are presented for random potentials but naturally extend to
random torques, as shown in SI~\cite{supp}.

\emph{Numerical simulations}.  We start by presenting results from
numerical simulations of $N$ run-and-tumble particles (RTPs) with
excluded volume interactions on a two-dimensional
lattice~\cite{thompson_lattice_2011,soto2014run,sepulveda2016coarsening,kourbane-houssene_exact_2018,whitelam_phase_2018}
of size $L\times L$ and periodic boundary conditions. Each particle
has an orientation $\hat{e}_\theta = (\cos{\theta}, \sin{\theta})$
with $\theta \in [0,2\pi)$, and reorients to a new random direction
with rate $\alpha$. In the absence of disorder, activity is accounted
for by hops from the position $\vec{i}$ of a particle to any
neighboring site $\vec{j} = \vec{i} + \hat{u}$ with rate
$W_{ \vec{i}, \vec{j} } = \max \left[ v \hat{u} \cdot \hat{e}_\theta
  ,0 \right]$, where $v$ controls the propulsion speed.  Interactions
between the particles are accounted for by modifying the hoping rates
according to
$W^\mathrm{int}_{ \vec{i}, \vec{j} } = W_{ \vec{i}, \vec{j} } (1 -
n_{\vec{j}} / n_{M})$ with $n_{\vec{j}}$ the number of particles at
$\vec{j}$ and $n_{M}$ the maximal occupancy. For large enough
$v/\alpha$ and densities, as shown in Fig.~\ref{Fig1}(a), the system
displays MIPS~\cite{thompson_lattice_2011}.  The quenched disorder is
modeled using
$W_{ \vec{i}, \vec{j} } = \max \left[ v \hat{u} \cdot \hat{e}_\theta -
  (V_{\vec{j}} - V_{\vec{i}}) ,0 \right]$ with $V_{\vec{i}}$ a random
potential drawn from a {\it bounded} uniform distribution,
$V_{\vec i} \in [-\Delta V,\Delta V]$. Here, the lattice spacing and
the mobility are set to one.

Surprisingly, Fig.~\ref{Fig1}(b) suggests that the phase separation is
washed out by the random potential. The resulting disordered phase
displays, however, a non-trivial structure, suggestive of interesting
correlations. We quantify the latter using the pair correlation
function
$g(r)=\overline{\frac{1}{L^2}\sum_j \langle n_{\vec j}n_{\vec j+\vec
    r} \rangle}$ where $r \equiv |{\vec r}|$, the brackets represent a
steady-state average, and the overline an average over disorder
realisations. In the absence of disorder, phase separation translates
into a linear decay of $g(r)$, as shown in Fig.~\ref{Fig1}(c). On the
contrary, in the presence of disorder, the correlations are found to
decay logarithmically, $g(r) \sim \log(L/r)$, as shown in
Fig.~\ref{Fig1}(d). This corresponds to a structure factor
$S(q)\sim q^{-2}$ for small $q$.

To explain the disappearance of phase separation and the emergence of
non-trivial correlations, we first introduce a phenomenological model
which captures the latter in a dilute system. This then suggests a
field-theoretic perspective which predicts the existence of weak- and
strong-disorder regimes. It first allows us to characterize the
disorder-induced persistent currents that flow in the system and then
to come back to the arrest of MIPS. Using the field theory, we
identify the lower critical dimension as $d_c=4$ and estimate the
Imry-Ma length scale above which phase separation is arrested in $d<4$.

\emph{Phenomenological model for a dilute system}. Random potential
and torques impact many aspects of the single-particle dynamics. As we
show, all the emerging phenomenology reported here can be traced back
to a single aspect: the emergence of ratchet currents. When a
localized asymmetric potential centered around ${\bf r}_0$ is placed
in an active fluid of non-interacting RTPs, the stationary density
field $\langle \rho({\bf r}) \rangle$ in the far field of the
potential is~\cite{baek_generic_2018}
\begin{equation}
 \langle \rho({\bf r}) \rangle = \rho_0 +  \frac{\beta_\mathrm{eff}}{S_d}  \frac{({\bf r} - {\bf r}_0) \cdot {\bf p}}{|{\bf r} - {\bf r}_0|^d} + {\cal O} \left( |{\bf r} - {\bf r}_0|^{-d} \right). 
\label{eq:dipole}
\end{equation}
Here, $S_d = 2 \pi^{d/2} / \Gamma(d/2)$, $\rho_0$ is the density of
the active fluid, $\beta_\mathrm{eff} \equiv 2 \alpha /v^2$, and the
mobility of the particles is set to one. The vector ${\bf p}$ is given by the
average force exerted by the potential on the active fluid and is thus
proportional to the overall density. In the presence of torques,
Eq.~\eqref{eq:dipole} still holds but with a renormalized
${\bf p}$~\cite{supp}. Given the analogy between Eq.~\eqref{eq:dipole}
and electrostatics, we follow Ref.~\cite{baek_generic_2018} and refer
to the force ${\bf p}$ as a dipole.
\begin{figure} [t]
	\center
	\includegraphics[width=1.0\linewidth]{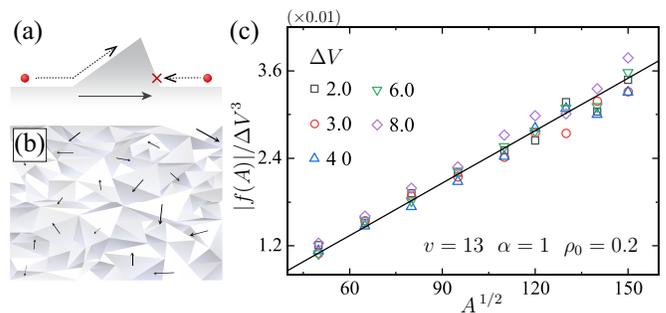}
	\caption{ {\bf (a)} In $d=1$, asymmetric potentials leads to a non-zero average force on the particles, indicated by a bold arrow. This is accompanied by a non-vanishing ratchet current. {\bf (b)} In $d=2$, a random potential  leads to a steady-state field of random forces exerted on the particles. {\bf (c)} Measurement of the force, $f(A)$, exerted on the active particles inside a region of area $A$ in the presence of a random potential. The amplitude of the force is quantified by  $|f(A)| \equiv ( \overline{f^2 (A)} )^{1/2}$, where $f(A)$ is obtained by  time-averaging $\sum_{\vec{i} \in A} n_{\vec{i}} (V_{\vec{i}+\vec{e}} - V_{\vec{i}-\vec{e}})/2$ with $\vec{e}$ an arbitrary unit vector.)
}
	\label{Fig2}
\end{figure}  

With Eq.~\eqref{eq:dipole} in mind, we consider a phenomenological
model in which the bounded random potential is modelled as a
superposition of random independent dipoles. Each dipole exerts a
force on the active particles in a direction dictated by the local
potential, as sketched in Figs.~\ref{Fig2}(a) and (b).  To test this
random dipole picture numerically, we measure in Fig.~\ref{Fig2}(c)
the force $f(A)$ exerted along an arbitrary direction by the random
potential on the particles inside an area $A$. Consistent with our
phenomenological model, $f(A)$ scales as $\sqrt{A}$. This should be
contrasted with equilibrium systems, where $f(A)$ is expected to scale
as $A^{1/4}$. Indeed, a random bounded potential $V({\bf r})$ leads,
in equilibrium, to a force density
$\propto \beta^{-1} \rho_0 \nabla \exp(-\beta V)$. Integrating over an
area $A$ solely leads to a boundary contribution proportional to
$\int_{\partial A} \exp(-\beta V) \vec n d\ell$. Only the fluctuations
of $V({\bf r})$ contribute, leading to the $A^{1/4}$ scaling.
Figure~\ref{Fig2}(c) thus highlights the non-equilibrium nature of the
system: the ratchet effect induced by the random potential leads to an
emerging force field with short-range correlations. Finally, the
scaling of $F(A)$ as $\Delta V^3$ in this dilute regime is consistent
with a perturbative result which predicts $|{\bf p}| \sim \Delta V^3$
as $\Delta V \to 0$~\cite{baek_generic_2018}, despite the relatively
large values of $\Delta V$ used here.

We now use the phenomenological model to predict the structure factor based on the random dipole picture. The  dipole density field ${\bf P}({\bf r})$ is randomly drawn from a distribution such that the spatial components of ${\bf P}$ satisfy $\overline{P_i ({\bf r})} = 0$ and $\overline{P_i ({\bf r}) P_j ({\bf r}')} = \chi^2 \delta_{ij} \delta^d({\bf r} - {\bf r}')$, notably lacking spatial correlations in ${\bf P}({\bf r})$. 
Denoting $\langle \phi({\bf r}) \rangle \equiv \langle \rho({\bf r})\rangle - \rho_0$, a direct computation, detailed in~\cite{supp}, leads from Eq.~\eqref{eq:dipole} to the  disorder-averaged structure factor:
\begin{equation}
{S}({\bf q}) \equiv \overline{\langle \phi({\bf q})  \phi(-{\bf q})\rangle} =  \frac{\beta_\mathrm{eff}^2 \chi^2}{q^2}~,
\label{eq:structure_factor}
\end{equation}
with $q \equiv |{\bf q}|$. Note that, in the dilute (noninteracting)
regime, the computation simplifies thanks to ${S}({\bf
  q})=\overline{\langle \phi({\bf q}) \rangle \langle\phi(-{\bf
    q})\rangle}$. Including interactions between the particles is
possible at the level of
Eq.~\eqref{eq:dipole}~\cite{granek_bodies_2020}, which would only
change the prefactor of $q^{-2}$ in
Eq.~\eqref{eq:structure_factor}. We stress that these predictions,
illustrated for scalar active matter, should hold for many active
systems, including polar and nematic ones, in the disordered
phase. In~\cite{supp} we indeed report long-range correlations in a
dilute system with aligning interactions.

Remarkably, the functional form of $S({\bf q})$ predicted by the
phenomenological model agrees well with our numerics even when the
particle density is not small, see Fig.~\ref{Fig1}(d). Building on
the successful predictions of the phenomenological model, we now
propose a field-theoretical description of scalar active matter
subject to quenched random potentials.

\emph{Field-theoretic treatment}.  Our results suggest that the
overall density is homogeneous at large scales, with small
fluctuations, so that the system can be described by a linear field
theory. This assumption will be checked in
the Section \textit{Strong disorder regime} 
using a self-consistency criterion. To model the force field emerging
from the ratchet effect due to the bounded random potential $V({\bf
  r})$, we introduce a quenched {\it random force} density ${\bf
  f}({\bf r})$ acting on the active fluid. We consider the dynamics:
\begin{eqnarray}
\label{Eq:phi} \frac{\partial}{\partial t} \phi({\bf r}, t) &=& - \nabla \cdot {\bf j}({\bf r}, t)~, \\
\label{Eq:j} {\bf j} ({\bf r}, t) &=& -  \nabla \mu [\phi] +  {\bf f}({\bf r}) + \sqrt{2D} \boldsymbol{\eta} ({\bf r}, t)~,
\end{eqnarray}
where $\phi({\bf r},t)$ denotes density fluctuations, ${\bf j}({\bf
  r}, t)$ is the linearized current, and $\boldsymbol{\eta} ({\bf r},
t)$ is a unit Gaussian white noise field. The mobility is set to one
and the random-force density satisfies $\overline{f_i ({\bf r})} = 0$
and $\overline{f_i ({\bf r}) f_j ({\bf r}')} = \sigma^2 \delta_{ij}
\delta^d({\bf r} - {\bf r}')$. To linear order in $\phi$ we set
\begin{equation} \label{Eq:mu}
\mu [\phi({\bf r}, t)] =  u \phi({\bf r}, t) - K \nabla^2 \phi({\bf r}, t)~,
\end{equation} 
with $u, K >0$ to ensure stability.  Note that, in the
small-fluctuation regime, $\boldsymbol{\eta}$ and ${\bf f}$ are
independent of $\phi$, while $\sigma$, $D$, $u$ and $K$ generically
depend on the mean density $\rho_0$.  Much work has been done, in
other contexts, on a single particle subject to a random
force~\cite{fisher_random_1984,bouchaud1990classical}. Our results
complement these classical works at the level of collective
modes. The structure factor is then given by~\cite{supp}:
\begin{equation} \label{Eq:str_phi}
 S({\bf q}) = \frac{  \sigma^2}{  q^2 (u + Kq^2)^2} + \frac{D}{(u + K q^2)}~.
\end{equation}
 Note that the small $q$ behavior of the structure factor reproduces
 the scaling $S(q) \propto q^{-2}$ predicted by the phenomenological
 model and observed in the numerics of Fig. \ref{Fig1}.  In fact,
 comparing Eqs.~\eqref{eq:structure_factor} and \eqref{Eq:str_phi}
 shows that $\sigma/u$ is proportional, in the dilute regime, to the
 inverse effective temperature: $\sigma/u \propto \chi
 \beta_\mathrm{eff}$~\footnote{With interactions, one finds
 $\chi/P'(\rho_0) = \sigma/u$, with $P$ the pressure of the active
 fluid~\cite{solon_pressure_2015-1,granek_bodies_2020}}. Interestingly,
 noise and interactions are subleading as $q \to 0$.

To further understand this result, we use a Helmholtz-Hodge
decomposition of the random force field: ${\bf f}({\bf r}) = -\nabla
U({\bf r}) + \boldsymbol{\xi} ({\bf r})$. Here $U({\bf r})$ is an {\it
  effective potential} reconstructed from the random force. Its
statistical properties, as we show below, are very different from
those of the potential $V({\bf r})$ which is short-range
correlated. The reconstructed vector field $\boldsymbol{\xi} ({\bf
  r})$ satisfies $\nabla \cdot \boldsymbol{\xi}({\bf r}) = 0$, so that
it impacts the current ${\bf j}$ but does not influence the dynamics
of the density field. To enforce the delta correlations of ${\bf
  f}({\bf r})$ together with its statistics, we set $\overline{ U({\bf
    q}) U ({\bf q}') } = {\sigma^2}{q^{-2}} \delta^d_{ {\bf q}, -{\bf
    q}' }$, $\overline{ \xi_i ({\bf q}) \xi_j ({\bf q}') } = \sigma^2
\left( \delta_{ij} - {q_i q_j'}/{q^{2}} \right) \delta^d_{ {\bf q},
  -{\bf q}' }$,
and $\overline{ U({\bf q}) \boldsymbol{\xi} ({\bf q}') } = 0$ so that
$U$ and ${\bf \xi}$ are intimately related.  Inserting the
decomposition into Eqs.~\eqref{Eq:phi} and \eqref{Eq:j} shows that the
density fluctuations of active particles in the disordered setting
behave as those of passive particles in an effective potential $U({\bf
  r})$.  The statistics of $U({\bf r})$ are those of a Gaussian
surface~\footnote{$U({\bf r})$ can be constructed from ${\bf f}({\bf
  r})$ by solving $\nabla^2 U({\bf r}) = \nabla \cdot {\bf f}({\bf
  r})$.}---a self-affine fractal with deep wells.  This effective
potential captures the component of the non-equilibrium current ${\bf
  j}$ which accounts for both the clustering and the long-range
correlations observed in our numerics. (See Fig.~\ref{Fig1}(b) and
Supplementary Movies, which show how MIPS is destabilized by the
introduction of the random potential.)

\begin{figure} [t!]
	\center
	\includegraphics[width=1.0\linewidth]{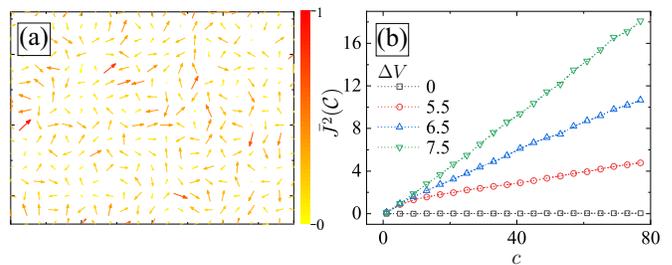}
	\caption{
	The current induced by disorder and its statistical properties. {\bf (a)} Current vector map for a realization of disorder. The color code is the steady-state current normalized by the maximum value measured.  {\bf (b)} The variance of the sum of current $J({\cal C})$ along a contour ${\cal C}$ as a function of the its perimeter $c$. 
	Parameters: $v=13$, $\alpha = 1$, $\Delta V = 6.5$ in (a).
	}
	\label{Fig_current}
\end{figure}

\emph{Persistent currents.}  The random force density ${\bf f}({\bf
  r})$ is a nonconservative vector field due to the divergence-free
part $\boldsymbol{\xi}({\bf r})$. While this term does not influence
the density field, it induces currents in the system. To quantify
them, we consider a closed contour ${\cal C}$. Taking the curl of the
total current and averaging over noise, one finds $\langle \nabla
\times {\bf j} \rangle = \nabla \times {\boldsymbol\xi} $.
Integrating this relation over a domain enclosed by ${\cal C}$, we
obtain, using Stokes theorem, that the circulation of $\langle {\bf
  j}\rangle$ is entirely controlled by ${\bf \boldsymbol\xi}$:
$J({\cal C}) \equiv \oint_{\cal C} \mathrm{d} {\bf l} \cdot {\bf
  j}({\bf r})= \oint_{\cal C} \mathrm{d} {\bf l} \cdot
{\boldsymbol\xi}({\bf r})$. $J({\cal C})$ is thus a sum of
uncorrelated random numbers and we predict its variance to scale as
the perimeter of the contour ${\cal C}$, with a slope proportional to
the disorder strength $\sigma$. This is confirmed by our numerics in
Fig.~\ref{Fig_current}(b). Furthermore, the steady-state current
induced by a realization of the random potential is shown in
Fig.~\ref{Fig_current}(a). It should be constrasted with the
equilibrium case in which currents vanish in the steady-state.

\emph{Strong-disorder regime and self-consistency of the linear theory}.
The linear theory used in the previous section is valid as long as density fluctuations are small compared to the mean density. To detect a possible departure from this scenario, we measure the  density fluctuations across a length $\ell$ through $\overline{ \langle \delta \rho^2(\ell) \rangle } = 2 \left[ g(a) - g(\ell) \right]$, with $a$ a short-distance cutoff. The fluctuations have to remain small compared to the natural scales of the density field:  $\overline{ \langle \delta \rho^2(\ell) \rangle }  \ll \rho_b^2$ with  $\rho_b \equiv \min(\rho_0, \rho_M - \rho_0)$. Here $\rho_0$ and $\rho_M$ are the average and maximal particle densities. Using Eq.~\eqref{Eq:str_phi}, we find for large $\ell$
\begin{equation}
	\frac{\overline{ \langle \delta \rho^2(\ell) \rangle }}{\rho_b^2} = \begin{dcases}
	\frac{ \sigma^2  \ln (\ell/a)}{\pi u^2 \rho_b^2}  & \mathrm{for} ~~ d = 2\\
	\frac{ \sigma^2  a^{2-d}}{(d-2) S_d u^2 \rho_b^2}  & \mathrm{for} ~~ d > 2~.
	\end{dcases}
\end{equation}
For $d>2$, the linear theory thus holds if $ \sigma \ll u \rho_b
\sqrt{(d-2) S_d a^{d-2}}$, namely, whenever the disorder is weak
enough. For strong enough disorder, the breakdown of the linear
approximation indicates the possibility of a different behavior for
$S(q)$. For $d=2$, the criterion is valid only for length scales
satisfying $\ell \ll \ell^\ast$ with $\ell^\ast \equiv a \exp(\pi u^2
\rho_b^2 / \sigma^2 )$. Note that this length scale is exponential in
the square of the ratio between the effective temperature and the
disorder strength, since $\sigma/u \propto \beta_\mathrm{eff}
\chi$. This can be estimated using Fig.~\ref{Fig2}, leading to very
large length scales, well beyond the reach of our numerics.  \if{This
  is likely due to a combination of the followings. For high or low
  densities, since the density is constrained, we expect $u$ to be
  very large. Furthermore, the disorder strength is a non-trivial
  function of the random potential strength. As expected and shown in
  the SI, it is bounded as a function of the variance of the potential
  and cannot take arbitrarily large values. This is expected since the
  effect of the potential is essentially unchanged when particles
  cannot cross it. Finally, by decreasing the effective temperature
  the ratchet effect is weakened and with it the value of
  $\sigma$.}\fi We suggest in the SI an alternative numerical approach
to study the strong-disorder regime in $d=2$ using passive particles
in a Gaussian surface. The resulting correlation function shows clear
deviations from the logarithmic behavior on large length scales.

\emph{Lower critical dimension.} Our linear theory offers an avenue to
test the stability of phase separation against weak disorder. To do
so, we note that the Helmholtz-Hodge decomposition implies that the
dynamics of $\phi({\bf r})$ are similar to an equilibrium dynamics in
a correlated random potential $U({\bf r})$.  This allows employing an
Imry-Ma argument~\cite{imry_random-field_1975,aharony_lowering_1976}
in order to obtain the lower critical dimension $d_c$ below which
phase separation is suppressed at large scales. To do so, we consider
a domain of linear size $\ell$. The surface energy of the domain is
given by $\gamma \ell^{d-1}$ with $\gamma$ the surface
tension~\footnote{Unlike equilibrium systems, $\gamma$, when defined
through momentum fluxes, can be negative for scalar active matter with
pairwise forces, leading to a bubbly liquid
phas~\cite{tjhung_cluster_2018,shi_self-organized_2020}. We expect our
result to hold provided that the cost of the domain wall in the
steady-state distribution increases as $\ell^{d-1}$ which is needed
for overall phase separation between dilute-gas and bubbly-liquid
phases.}.  On the other hand, the contribution of the disorder to the
energy of the domain is, to leading order, $E(\ell) = \int_{\ell^d} \mathrm{d}^d {\bf r}'
~ \rho_0 U({\bf r}')$. The typical energy of a domain of size $\ell$
is thus given by $\sqrt{\overline{ E(\ell)^2 }} = \sigma \rho_0
\ell^{(d+2)/2}$. Comparing the two energy scales shows the lower
critical dimension to be $d_c=4$.  In lower dimensions the
contribution of the surface energy is negligible on large enough
length scales and a system of size $L$ does not phase separate if $L
\gg \ell_{IM} \simeq \left[ {\gamma}/ ({\sigma} \rho_0 ) \right]^{\frac{2}{4-d}}$,
which we term the Imry-Ma length scale. Numerically, we indeed confirm that the coarsening to a single macroscopic domain is only observed for small system sizes. Correspondingly, a transition from linearly decaying to logarithmically decaying pair-correlation functions with increasing $L$ is reported in SI~\cite{supp}.

\begin{figure} [t]
	\center
	\includegraphics[width=1.0\linewidth]{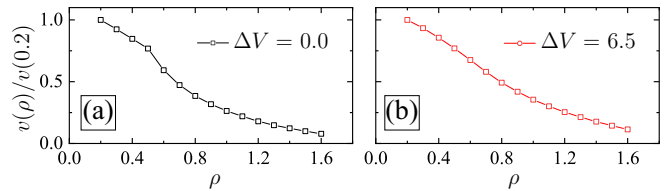}
	\caption{Measurement of the effective self-propulsion speed $v(\rho)$ in our simulations without {\bf (a)} and with {(\bf b)} disorder for $v=13$ and $\alpha=1$. Both systems exhibit a similar decay which, in the absence of disorder, would lead to MIPS. Note that the kink observed for $v(\rho)$ in the left panel stems from the occurrence of phase separation.
	}
	\label{Fig4}
\end{figure}

Note that the Imry-Ma argument rules out the existence of a macroscopic ordered/dense phase. Alternatively, the absence of MIPS could stem from the suppression of the feedback loop between a slowdown of particles at high density and their tendency to accumulate where they move slower. Reformulated as a 
mean-field theory, this feedback loop translates into an instability criteria for a homogeneous system of density $\rho$ whenever $\rho v'(\rho)<-  v(\rho)$~\cite{tailleur2008statistical,cates_motility-induced_2015}, where $v(\rho)$ is an effective propulsion speed in a system of density $\rho$. We report in Fig.~\ref{Fig4} the measurement of $v(\rho)$ for our system, defined as the mean hoping rate of particles along their orientation, with and without the random potential. Both systems show a similar decay which, at mean-field level, would predict the occurrence of MIPS. It is thus the non-trivial correlations induced by disorder that make MIPS disappear at large scales, despite an underlying instability at mean-field level. 
The disorder-induced disappearance of MIPS thus has a very different origin than its arrest by 
diffusiophoretic~\cite{pohl2014dynamic,zottl2014hydrodynamics} or hydrodynamic~\cite{matas2014hydrodynamic} interactions that directly prevent a kinetic hindrance at the microscopic scale.

\textit{Conclusion.} In this Letter we have shown how random quenched
potentials and torques lead to a non-trivial phase in scalar active
matter with anomalous correlations that prevent phase
separation. Interestingly, while the transition to collective motion
is more robust to disorder than the corresponding ferromagnetic
transition in
equilibrium~\cite{toner_hydrodynamic_2018,toner_swarming_2018}, the
converse holds for scalar phase separation: the lower critical
dimension is larger in the active case ($d_c=4$) than in the passive
one ($d_c=2$).  We also note a strong difference between the
one-dimensional active case, in which disorder promotes
clustering~\cite{ben_dor_ramifications_2019}, and the two-dimensional
one, in which MIPS is destroyed by disorder at large scale.  We expect
our results, presented here for RTPs, to hold for generic scalar
active systems, including active Brownian and active Ornstein
Uhlenbeck particles.  Experimentally, we expect our results to be
relevant for a large class of systems. Long-ranged correlations, forces
on obstacles, and current circulations could be tested using
self-propelled colloids~\cite{howse2007self,bricard2013emergence} and
shaken grains~\cite{deseigne2010collective} on irregular surfaces---at
least in their dilute, disordered phase---as well as swimming bacteria
in disordered
media~\cite{bhattacharjee_confinement_2019,bhattacharjee_bacterial_2019-1}. The
suppression of phase separation and MIPS-related physics could be
studied in experiments using self-propelled
colloids~\cite{buttinoni_dynamical_2013,van2019interrupted} or
bacteria~\cite{liu2019self,curatolo2020cooperative}. \if{In addition,
  our results also call for a more general study of the influence of
  disorder-induced long-range correlations on other active matter
  systems, for instance on active
  nematics~\cite{chardac_meandering_2020}.}\fi It would also be of
interest to see whether the bubbly phase, uncovered recently
in~\cite{tjhung_cluster_2018,shi_self-organized_2020}, exhibits
different behavior under disorder. Finally, we note that random
potentials lead to ratchets in many nonequilibrium systems, whether
classical or quantum, far beyond the realm of active matter. Since
these currents are the building blocks of our field-theoretical
treatment, we expect our results to play a role in many nonequilibrium
systems experiencing random potentials.

{\it Acknowledgments:}  SR, YK \& MK were supported by an NSF5-BSF grant (DMR-170828008). SR \& YK were also supported by the Israel Science Foundation and the National Research Foundation of Korea (2019R1A6A3A03033761). SR, JT \& YK acknowledge support from a joint CNRS-MOST grant.
 The authors benefited from participation in the 2020 KITP program on \textit{Active Matter} supported by the Grant NSF PHY-1748958.

 \bibliographystyle{apsrev4-1}
 \bibliography{Disorder_Bib}
\end{document}


\title{Supplemental Information of Manuscript ``Disorder-Induced Long-Ranged Correlations in Scalar Active Matter''}

\author{Sunghan Ro}
\affiliation{Department of Physics,
  Technion-Israel Institute of Technology,
  Haifa, 3200003, Israel.}
\author{Yariv Kafri}
\affiliation{Department of Physics,
  Technion-Israel Institute of Technology,
  Haifa, 3200003, Israel.}
\author{Mehran Kardar}
\affiliation{Department of Physics,
  Massachusetts Institute of Technology,
  Cambridge, Massachusetts 02139, USA.}
\author{Julien Tailleur}
\affiliation{Universit{\' e} de Paris, laboratoire Matière et Systèmes Complexes (MSC),
  UMR 7057 CNRS, 75205 Paris, France.
}




\maketitle


\tableofcontents{}

\section{Phenomenological model}
\subsection{Analytical computation of the structure factor}
Here we evaluate the structure factor based on the random dipole picture of the phenomenological model. 
The density $\phi ({\bf r})$ in the presence of the random dipole field ${\bf P}({\bf r})$ reads 
\begin{equation} \label{Eq:phi}
  \langle \phi ({\bf r}) \rangle = \beta_\mathrm{eff} \int \mathrm{d} {\bf r}' ~ {\bf P}({\bf r}') \cdot \nabla_{\bf r} G({\bf r} - {\bf r}')~,
\end{equation}
where $G({\bf r} - {\bf r}')$ is the Green function of the Poisson equation. Note that if we have a delta-distributed dipole ${\bf P}({\bf r}) = {\bf p} \delta^d({\bf r} - {\bf r}_0)$ in a $d$-dimensional free space, Eq.~\eqref{Eq:phi} reads 
\begin{equation}
  \langle \phi ({\bf r}) \rangle = \frac{\beta_\mathrm{eff}}{S_d} \frac{({\bf r} - {\bf r}_0) \cdot {\bf p}}{|{\bf r} - {\bf r}_0|^d}~,
\end{equation}
which recovers Eq.~(1) of the main text. 
In what follows we use the Fourier convention:
\begin{equation}
  \nonumber	f({\bf q}) = \frac{1}{\sqrt{V}} \int \mathrm{d}^d {\bf r} ~ e^{-i {\bf q} \cdot {\bf r}} f({\bf r}) ~~~~ \mathrm{and} ~~~~ f({\bf r}) = \frac{1}{\sqrt{V}} \sum_{\bf q} e^{i {\bf q} \cdot {\bf r}} f({\bf q})~,
\end{equation}
with $V$ denoting the volume of the system. In the Fourier representation, the convolution in Eq.~\eqref{Eq:phi} is written as a product 
\begin{equation}
  \nonumber	\langle \phi ({\bf q}) \rangle =  \beta_{\rm eff}V^{1/2} i {\bf q} \cdot {\bf P}({\bf q}) G({\bf q} )~.
\end{equation}
Here $G({\bf q}) = - V^{-1/2} q^{-2}$ with $q = |{\bf q}|$, and the first two moments of ${\bf P}({\bf q})$ reads
\begin{eqnarray}
  \nonumber	\overline{P_i ({\bf q})} &=& 0 ~,\\
  \nonumber	\overline{P_i ({\bf q}) P_j({\bf q}')} &=& \chi^2 \delta_{ij} \delta_{{\bf q}, -{\bf q}'}~\;.
\end{eqnarray}
Using these relations, the structure factor is calculated in the dilute regime where  $\overline{ \langle \phi ({\bf q})  \phi (-{\bf q}) \rangle }=\overline{ \langle \phi ({\bf q}) \rangle \langle \phi (-{\bf q}) \rangle }$ as
\begin{eqnarray}
  \nonumber	\overline{ \langle \phi ({\bf q})  \phi (-{\bf q}) \rangle } &=& \frac{\beta^2_\mathrm{eff}}{ q^4} \sum_{i,j=1}^d q_i q_j \overline{ P_i ({\bf q})  P_j (-{\bf q}) } \\
  \label{Eq:S}	&=& \frac{\beta^2_\mathrm{eff} \chi^2}{ q^2}~,
\end{eqnarray}
which is Eq.~(2) of the main text. 

\subsection{Long-ranged correlations in dilute, disordered phases}

\begin{figure} [t]
  \center
  \includegraphics[width=0.9\linewidth]{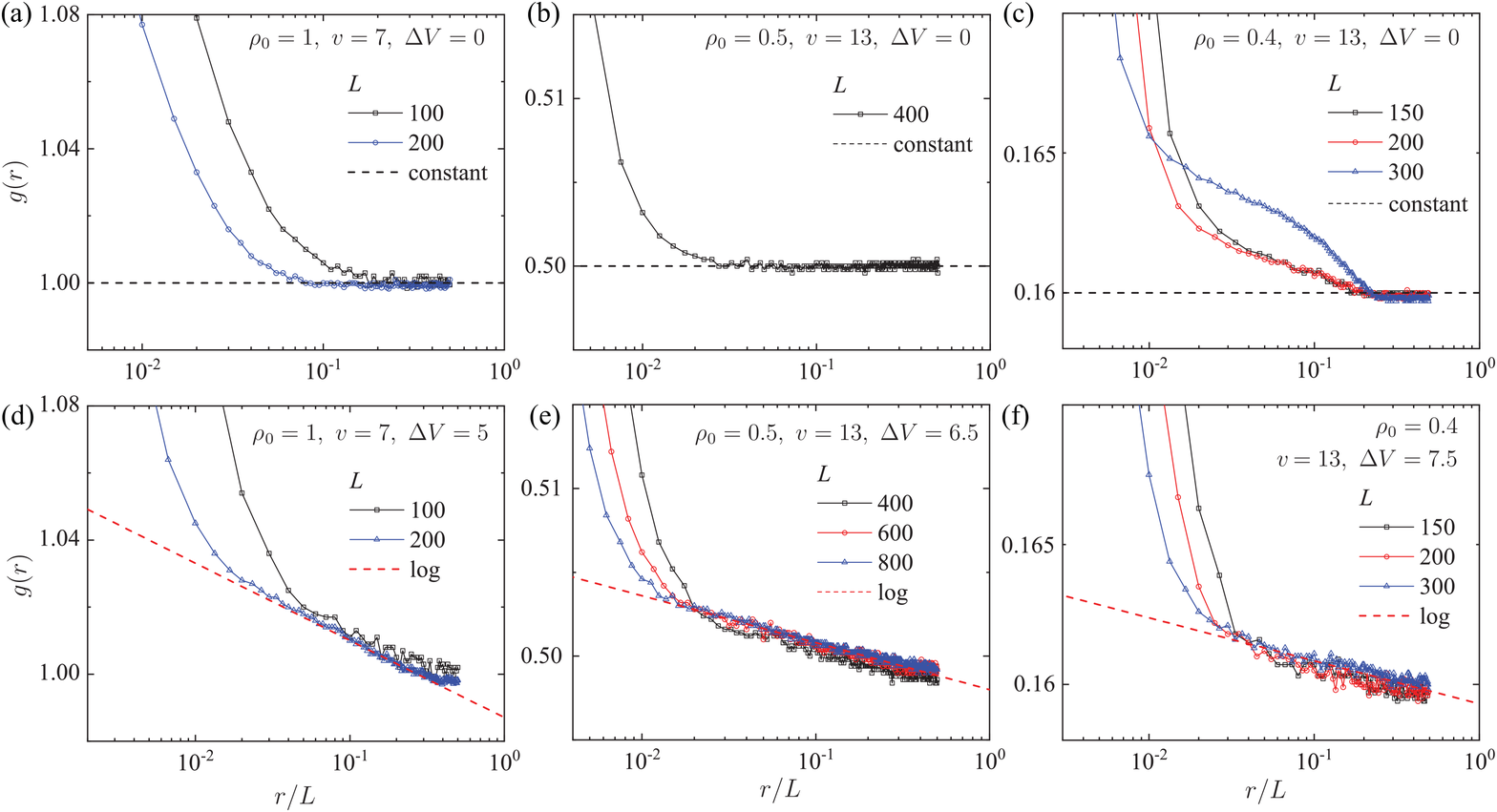}
  \caption{Density two-point correlation functions $g(r)$ obtained from weekly propelling or dilute active matter. 
    Upper panels: $g(r)$ obtained without disorder when (a) $v$ is small and (b) $\rho_0$ is small. Lower panels: $g(r)$ obtained with disorder when (c) $v$ is small and (d) $\rho_0$ is small. The panel (e) shows the correlation obtained from dilute active matter with the aligning interaction. Parameters: $\alpha=1$, $n_M = 2$, $\tau = 3.0$, $\theta_M = 0.1$. 
  }
  \label{Fig0}
\end{figure}

In Fig.~\ref{Fig0}, we show that the scale-invariant correlations are
also observed in dilute systems. We consider three cases. First, we
consider the model defined in the main text in the small $v$ or small
$\rho_0$ regimes, where MIPS is not observed. In the absence of
disorder, Fig.~\ref{Fig0}(a) and (b) show density-density correlations
$g(r)$ with short-ranged correlations. In the presence of disorder,
Figs.~\ref{Fig0}(c) and (d) show the long-ranged logarithmic
correlations predicted by Eq.~\eqref{Eq:S}. 

The third case we considered is the dilute, disordered phase of an
active polar system. To do so, we introduce aligning interactions in
the active lattice gas described in the main text. Specifically, given
a particle $a$ at site $\vec{i}$ with orientation $\hat{e}_\theta^a$,
we evaluate the local polarization of neighboring particles as
${\bf m}_{a;\vec{i}} = \sum_{b\neq a} \hat{e}_{\theta}^b$, where $b$
runs over the particles at the site $\vec{i}$ and its nearest
neighbors. In addition to hopping and uniform tumbles, the particles
may now undergo a change of direction resulting from the aligning
interactions. The corresponding rate  $Y_a$ is given by
\begin{equation}
  Y_a = \tau n_a \left(1 - \hat{m}_{a;\vec{i}} \cdot \{ \hat{e}_\theta \}_a \right) (1-\delta_{{\bf m}_{a;\vec{i}},0})~,
\end{equation}
where $\tau$ is the aligning interaction strength, $n_a$ is the number
of neighboring particles used to calculate ${\bf m}_{a;\vec{i}}$, and
$\hat{m}_{a;\vec{i}} = {\bf m}_{a;\vec{i}} / |{\bf
  m}_{a;\vec{i}}|$. When such an aligning step occurs, the particle
orientation change as $ \theta \to \theta + \varepsilon \Delta \theta$
where $\Delta \theta\in [0,\theta_M] $ is drawn uniformly at random,
and $\varepsilon=\pm 1$ is chosen to reduce the angle between
$\hat{e}_\theta^a$ and ${\bf m}_{a;\vec{i}}$. In Fig.~\ref{Fig0}(c),
we show that the system exhibits short-ranged correlations in its
dilute, disordered phase. Note that aligning interactions make the
correlation length larger than in the scalar system. In the presence
of disorder, we again observe the predicted long-ranged
correlations. This result confirms that, as expected, polar active
systems exhibit the same structure factor as scalar active systems in
their disordered phase. In experimental systems, the presence of
aligning interactions will thus not be an obstacle to measuring
long-ranged correlations as long as the systems are disordered.

\section{Field-theoretic treatment}
Here we calculate the structure factor within the linear field-theory presented in the main text.
Namely, the structure factor corresponding to
\begin{eqnarray}
  \label{Eq:MBphi} \frac{\partial}{\partial t} \phi ({\bf r}, t) &=& - \nabla \cdot {\bf j} ({\bf r}, t), \\
  \label{Eq:MBj} {\bf j} ({\bf r}, t) &=& - \nabla \mu [\phi] + {\bf f} ({\bf r}) + \boldsymbol{\eta} ({\bf r}, t).
\end{eqnarray}
with
\begin{equation}
  \nonumber	\mu [\phi({\bf r}, t)] = u \phi({\bf r}, t) - K \nabla^2 \phi({\bf r}, t)~\;.
\end{equation}
Here $\overline{f_i ({\bf r})} = 0$ and $\overline{f_i ({\bf r}) f_j ({\bf r}')} = \sigma^2 \delta_{ij} \delta^d({\bf r} - {\bf r}')$. The Gaussian white noise $\boldsymbol{\eta} ({\bf r}, t)$ is characterized by zero mean and $\langle \eta_i ({\bf r}, t) \eta_j ({\bf r}', t') \rangle = 2 D \delta_{ij} \delta^d({\bf r} - {\bf r}') \delta (t-t')$.

Writing Eq.~\eqref{Eq:MBphi} in Fourier space, we find
\begin{equation}
  \nonumber	\frac{\partial}{\partial t} \phi ({\bf q}, t) = - q^2 (u + K q^2) \phi({\bf q}, t) - i {\bf q} \cdot {\bf f} ({\bf q}) - i {\bf q} \cdot \boldsymbol{\eta} ({\bf r}, t)~.
\end{equation}
Solving this equation we obtain
\begin{equation}
  \nonumber	 \phi ({\bf q}, t) = \phi ({\bf q}, 0) e^{-\kappa ({\bf q}) t} - \frac{i {\bf q} \cdot {\bf f}({\bf q})}{\kappa ({\bf q})} \left( 1 - e^{-\kappa({\bf q})t} \right) - \int_0^t \mathrm{d} t' ~ e^{-\kappa({\bf q}) (t - t')}  i {\bf q} \cdot \boldsymbol{\eta} ({\bf r}, t)~.
\end{equation}
Here $\kappa ({\bf q}) \equiv q^2 (u + K q^2)$. Using this solution, we calculate the structure factor in the stationary state $S({\bf q}) = \lim_{t \rightarrow \infty} \overline{ \langle \phi ({\bf q}, t) \phi (-{\bf q},t) \rangle }$. This leads to
\begin{eqnarray}
  \nonumber	S({\bf q}) &=& \frac{q^2 \sigma^2}{\kappa({\bf q})\kappa(-{\bf q})} + \frac{2 D q^2}{ \kappa({\bf q})+\kappa(-{\bf q})}  \\
  \nonumber	&=& \frac{\sigma^2}{q^2 (u + Kq^2)^2} + \frac{D}{(u + Kq^2)}~,
\end{eqnarray}
where we used the fact that the second moments of the random variables in the Fourier representation read $\overline{f_i ({\bf q}) f_j ({\bf q}')} = \sigma^2 \delta_{ij} \delta^d_{{\bf q} , -{\bf q}'}$ and $\langle \eta_i ({\bf q}, t) \eta_j ({\bf q}', t') \rangle = 2 D \delta_{ij} \delta^d_{{\bf q} , -{\bf q}'} \delta ( t - t')$.

\begin{figure} [t]
  \center
  \includegraphics[width=0.8\linewidth]{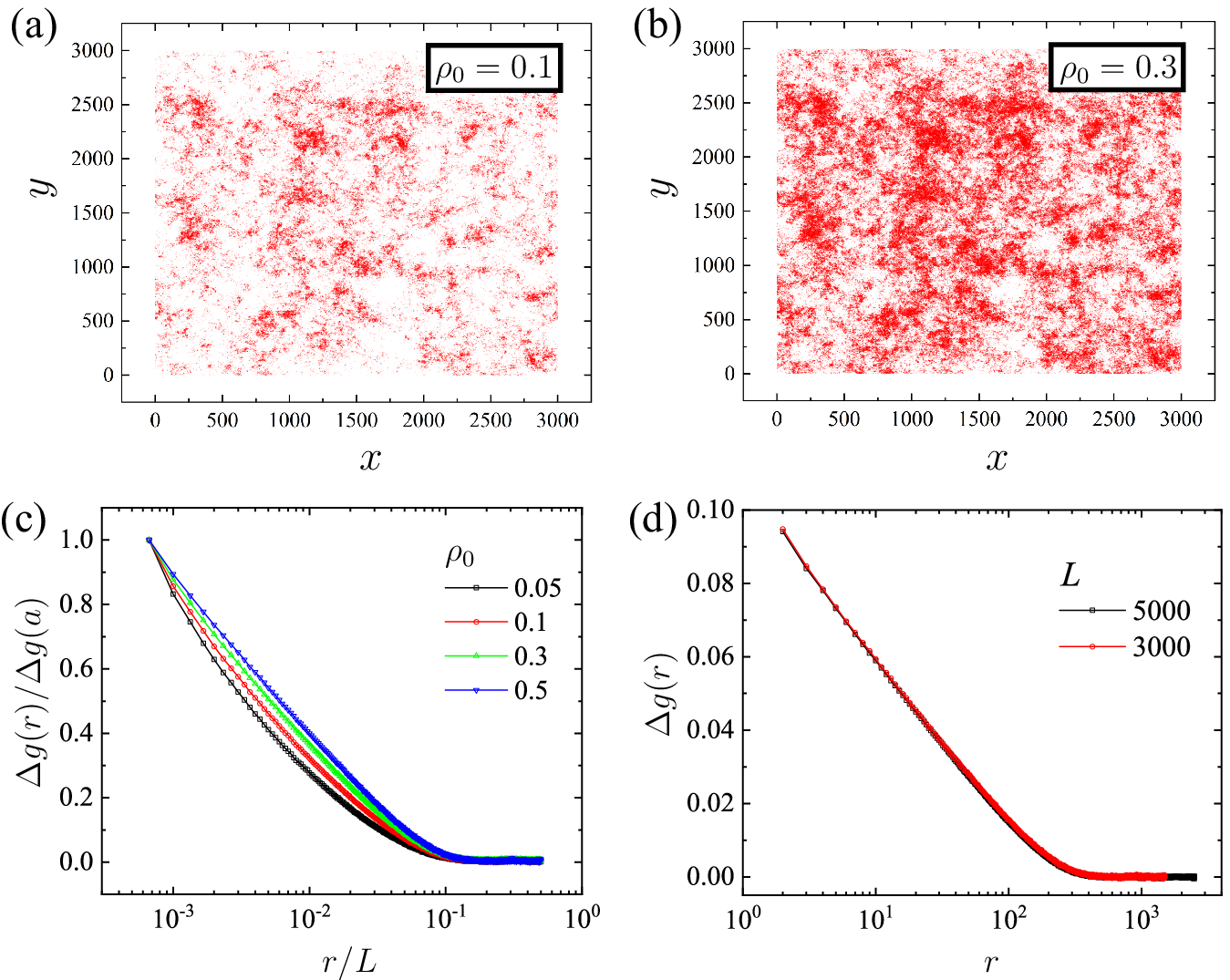}
  \caption{Distribution of particles with an average density $\rho_0=0.1$ (a) and $\rho_0=0.3$ (b) on a Gaussian surface. The plateaux exhibited by the pair correlation functions, which we collapse at small and large lengths, in (c) show a clear departure from the logarithmic decay predicted in the weak disorder regime. Note that the crossover takes place at a length independent from system size, as shown in panel (d) by overlapping data for $L=3000$ and $5000$.}
  \label{Fig1}
\end{figure}

\section{Strong-disorder regime} 

The strong-disorder regime proved beyond the numerical reach
of our lattice-based model. Nevertheless, we can use the
field-theoretical model to propose a heuristic approach to
study the pair correlation function. Note that this approach
will thus ignore any non-equilibrium non-linear contributions
that the field theory does not capture. Since the effective
potential exhibit deep wells, which are dominant for
determining the particle distribution, we expect these
non-linear contributions to play a less significant role and
to leave the results unaltered at a qualitative level.

To proceed further, we generate a random effective potential $U({\bf r})$ that satisfies the statistics of the Gaussian surface. Then, we introduce particles interacting with excluded volume interaction. Given the diverging depth of the potential well, temperature is not expected to play a large role and we consider the system at $T=0$. To do so, we simply fill up the system from the bottom of its deepest minima. This could be thought of as filling the Fermi-sea of a 
two-dimensional
Fermionic system in a Gaussian surface potential. Finally, we measure the pair correlation function and average over disorder realizations. 

In Fig.~\ref{Fig1}(a) and \ref{Fig1}(b), we show configurations obtained from the process described above. Note that the particle distribution is superficially similar to the configurations obtained from the simulations of run-and-tumble particles presented in Fig.~1(b) of the main text. 
In Fig.~\ref{Fig1}(c) and (d), we present the pair correlation functions, which we rescale as $\tilde g(r)\equiv \Delta g(r) / \Delta g(a)$ where $a$ is the lattice spacing $a$ and $\Delta g(r) \equiv g(r) - g(L/2)$. 
As predicted, when $r/L$ is small, $\tilde g(r)$ shows a logarithmic decay as a function of $r/L$. For large $r/L$, however, the pair correlation function deviates from this logarithmic behavior, consistently with the prediction of a  strong-disorder regime. 
	In Fig.~\ref{Fig1}(d), we show the crossover to be independent of system size, as predicted by our theory.
	
	\section{Imry-Ma length scale} 
	
\begin{figure} [h]
		\center
		\includegraphics[width=0.9\linewidth]{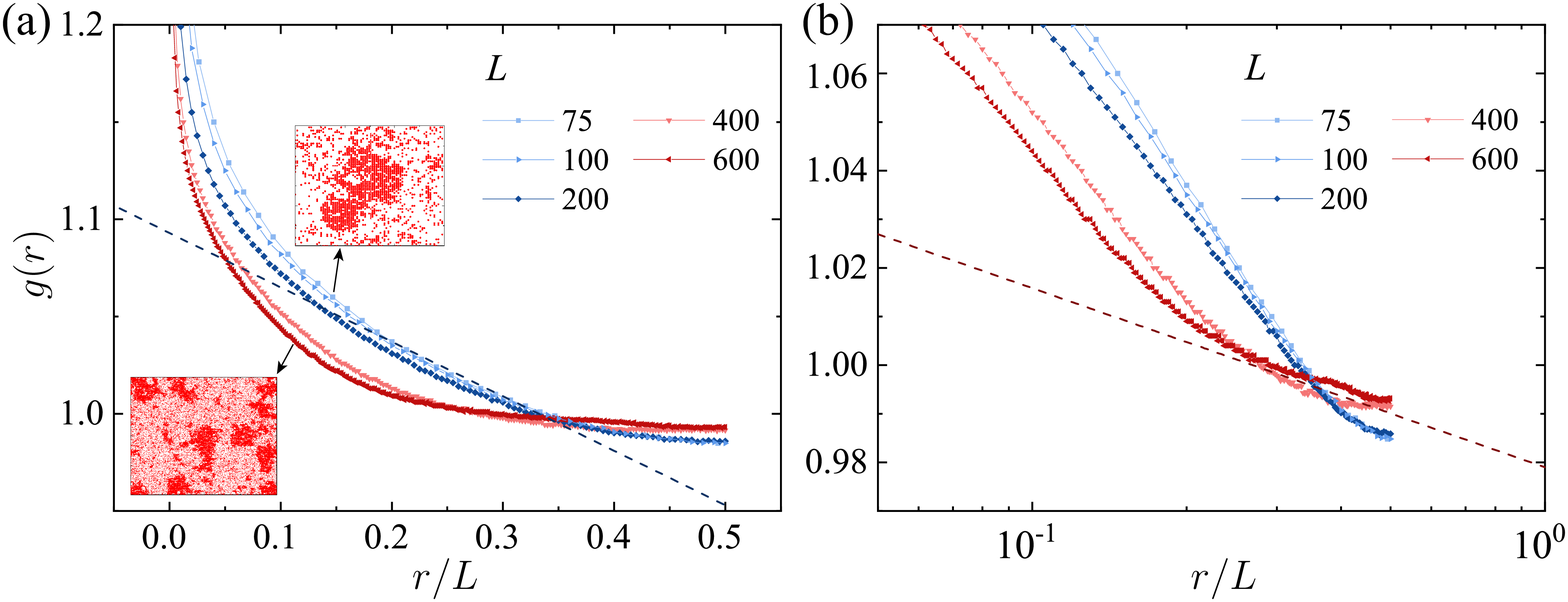}
		\caption{The density pair correlation function measured for $\Delta V = 2.5$ and $v/\alpha = 10$ for different system sizes. The data in blue corresponds to small system sizes and in red to larger systems. Two different behaviors are observed consistent with the existence of an Imry-Ma length scale. }
		\label{Fig2}
	\end{figure}

	In the following, we explore how the system transitions from phase-separated to homogeneous at large scales as the system size crosses the Imry-Ma length scale. To do so, we measure the pair-correlation function for different system sizes in the presence of disorder. We show an example of such a measurement in Fig. \ref{Fig2}(a) for a disorder strength $\Delta V = 2.5$. As shown in this Figure, there are two distinct regimes which depend on the system size. The data are correspondingly
	marked in blue and red. The crossover length between the two behaviors provides an approximate measurement of $\ell_{\rm IM}$. Indeed, for $L \ll \ell_\mathrm{IM}$: $g(r)$ decays linearly with $r/L$ as expected for a phase-separated system (see blue curves and black dashed line). In contrast, for  $L \gg \ell_\mathrm{IM}$, $g(r)$ decays logarithmically with $r/L$ as predicted in the main text and confirmed in  Fig. \ref{Fig2}(b) (see red curves and red dashed line). Sample configurations in each of the regimes are also shown as insets in  Fig. \ref{Fig2}(a). 

	\section{The Influence of Torques}
	
	Here we consider the effect of torques acting on a dilute gas of ABPs or RTPs. In particular, we show that a localized torque influences the far field and current densities in the same manner as an asymmetric potential. The only difference is that the dipole strength is no longer simply related to the force exerted by the asymmetric potential on the active fluid. This result implies that the phenomenological model and the field theoretical treatments of the main text remain unchanged in the presence of torques. This statement is supported by numerical simulations in two dimensions which include interactions between the particles and confirm the expected decay of the correlation function. 
	
	To proceed, we follow Ref.~[51] of the main text to derive the equation for the density field in the far-field of a localized torque. For completeness, we also account for a localized potential in the derivation. It should be clear to the reader that allowing only torques does not modify the results. In addition, for simplicity, the derivation is carried out in two-dimensions. The extension to higher dimensions is straightforward. 
	
	Consider the two-dimensional Fokker-Planck equation:
	\begin{equation} \label{Eq:FP}
	\partial_t P = - \nabla \cdot \left[ \left( v {\bf e}_\theta  -\nabla V({\bf r}) - D_t \nabla  \right) P \right] - \alpha P + \frac{\alpha}{2 \pi} \int \mathrm{d} \theta ~ P - \partial_\theta \left[ \Gamma ({\bf r}, \theta) P \right] + D_r \partial_\theta^2 P~.
	\end{equation}
	Here, $P({\bf r}, \theta)$ is the probability density of
        finding the particle at position ${\bf r}$ with an orientation
        ${\theta}$, $v$ is the particle's speed, ${\bf e}_\theta$ is
        an unit vector in the $\theta$ direction, $V({\bf r})$ is the
        potential, $D_{t(r)}$ is a translational (rotational)
        diffusion constant, and $\alpha$ is a tumbling rate. As in the
        main text, we set the mobilities to one. The localized torque
        exerted on the particle is accounted for through $\Gamma({\bf
          r}, \theta)$. We assume that $V({\bf r})$ and $\Gamma({\bf
          r}, \theta)$ are non-zero in the same region of space.
	
	Next, we introduce the marginal distributions for the probability density:
	\begin{equation}
	{\bf m}^{(n)} ( {\bf r} ) \equiv \int \mathrm{d} \theta ~ P({\bf r}, \theta) {\bf e}_\theta^{(n)}
	\end{equation}
	with 
	\begin{equation}
	{\bf e}_\theta^{(n)} \equiv \left[ \begin{array}{c}
	\cos (n \theta) \\ \sin (n \theta)
	\end{array} \right]~.
	\end{equation}
	Note that ${\bf m}^{(0)} ( {\bf r} ) \equiv [ \rho({\bf r}), 0]^\mathrm{T}$. We further define the marginal torque distributions
	\begin{eqnarray}
	\boldsymbol{\Gamma}^{(n)} ({\bf r}) &\equiv& \int \mathrm{d} \theta ~ \Gamma({\bf r}, \theta) P({\bf r}, \theta) \mathbb{C} {\bf e}_\theta^{(n)}
	\end{eqnarray}
	with
	\begin{equation}
	\mathbb{C} \equiv \left[ \begin{array}{cc}
	0 & -1 \\ 1 & 0
	\end{array} \right]~\;.
	\end{equation}
	Multiplying  Eq.~\eqref{Eq:FP} by  ${\bf e}_\theta^{(n)}$ and integrating over $\theta$, one finds that the equations for the marginal distributions are given in the steady-state by
	\begin{eqnarray}
	\label{Eq:0} 0 &=& - \nabla \cdot \left( v {\bf m}^{(1)} - \rho  \nabla V - D_t \nabla \rho \right)~, \\
	\label{Eq:n} 0 &=& -(\alpha + D_r n^2) \left( 1 - \mathbb{M}^{(n)} \right) {\bf m}^{(n)} - \frac{v}{2} \left( \mathbb{D} {\bf m}^{(n-1)} - \mathbb{D}^\dagger {\bf m}^{(n+1)} \right) + n \boldsymbol{\Gamma}^{(n)} ~\;.
	\end{eqnarray}
	Here, the operators $\mathbb{M}^{(n)}$, $\mathbb{D}$, and $\mathbb{D}^\dagger$ are defined as
	\begin{eqnarray}
	\mathbb{M}^{(n)} &\equiv& \frac{1}{\alpha + D_r n^2} \left[  \nabla \cdot (\nabla V) + D_t \nabla^2  \right]~, \\
	\mathbb{D} &\equiv& \left[ \begin{array}{cc}
	\partial_x & - \partial_y \\ \partial_y & \partial_x
	\end{array} \right]~, ~~~ \mathbb{D}^\dagger \equiv \left[ \begin{array}{cc}
	-\partial_x & - \partial_y \\ \partial_y & -\partial_x
	\end{array} \right]~.
	\end{eqnarray}
	
	Using Eq.~\eqref{Eq:n}, we can express  ${\bf m}^{(1)}$ and  ${\bf m}^{(2)}$ as
	\begin{eqnarray}
	\label{Eq:m1} {\bf m}^{(1)} &=& \mathbb{M}^{(1)} {\bf m}^{(1)} - \frac{l_r}{2} \left( \nabla \rho -  \mathbb{D}^\dagger {\bf m}^{(2)} \right) + \frac{1}{\alpha + D_r} \boldsymbol{\Gamma}^{(1)} \\
	\label{Eq:m2} {\bf m}^{(2)} &=& \mathbb{M}^{(2)} {\bf m}^{(2)} - \frac{l_r}{2} \frac{\alpha + D_r}{\alpha + 4 D_r} \left( \mathbb{D} {\bf m}^{(1)} -  \mathbb{D}^\dagger {\bf m}^{(3)} \right) + \frac{2}{\alpha + 4 D_r} \boldsymbol{\Gamma}^{(2)}~.
	\end{eqnarray}
	Here, $l_r \equiv v/(\alpha + D_r)$ is the run length.
	Expressing $- \nabla \cdot \left( v {\bf m}^{(1)} \right)$ in Eq.~\eqref{Eq:0} by combining Eqs.~\eqref{Eq:m1} and \eqref{Eq:m2}, one finds
	\begin{eqnarray}
	\nonumber - \nabla \cdot \left( v {\bf m}^{(1)} \right) &=&  -  l_r \sum_{a, b} \partial_a \partial_b \left[ (\partial_a V) \left( {\bf m}^{(1)} \cdot {\bf e}_b \right) \right] - D_t l_r \nabla^2 \nabla \cdot {\bf m}^{(1)}  \\
	\nonumber && + \frac{v l_r}{2} \nabla^2 \rho   - \frac{v l_r}{2} \nabla \cdot \mathbb{D}^\dagger {\bf m}^{(2)} - l_r \nabla \cdot \boldsymbol{\Gamma}^{(1)}~, \\
	\nonumber &=&  \frac{v l_r}{2} \nabla^2 \rho - l_r \nabla \cdot \boldsymbol{\Gamma}^{(1)} + l_r^2 \frac{\alpha + D_r}{\alpha + 4 D_r} \nabla \cdot \mathbb{D}^\dagger {\bf \Gamma}^{(2)} \\
	\label{Eq:vm1} && -  l_r  \sum_{a, b} \partial_a \partial_b \left[ (\partial_a V) \left( {\bf m}^{(1)} \cdot {\bf e}_b \right) \right] + \sum_{a,b,c} \partial_a \partial_b \partial_c \mathbb{H}_{abc}~, 
	\end{eqnarray}
	with 
	\begin{eqnarray}
	\nonumber \sum_{a,b,c} \partial_a \partial_b \partial_c \mathbb{H}_{abc} &=& - D_t l_r \nabla^2 \nabla \cdot {\bf m}^{(1)} - \frac{l_r^2}{2} \frac{\alpha + D_r}{\alpha + 4 D_r} \nabla \cdot \mathbb{D}^\dagger \sum_{a} \partial_a \left[  (\partial_a V) + D_t \partial_a \right] {\bf m}^{(2)} \\
	&& + \frac{v l_r^2}{4} \frac{\alpha + D_r}{\alpha + 4 D_r} \nabla \cdot \left[ \nabla^2 {\bf m}^{(1)} - (\mathbb{D}^\dagger)^2 {\bf m}^{(3)} \right]~.
	\end{eqnarray}
	Inserting Eq.~\eqref{Eq:vm1} into Eq.~\eqref{Eq:0}, the following two-dimensional Poisson equation is obtained
	\begin{eqnarray}
	\nonumber \nabla^2 \rho &=& - \beta_\mathrm{eff} \nabla \cdot (\rho \nabla V - l_r \boldsymbol{\Gamma}^{(1)}  ) \\
	\label{Eq:source} && + \beta_\mathrm{eff} \left[- l_r^2  \frac{\alpha + D_r}{\alpha + 4 D_r} \nabla \cdot \mathbb{D}^\dagger {\bf \Gamma}^{(2)} + l_r  \sum_{a, b} \partial_a \partial_b \left[ (\partial_a V) \left( {\bf m}^{(1)} \cdot {\bf e}_b \right) \right] -  \sum_{a,b,c} \partial_a \partial_b \partial_c \mathbb{H}_{abc} \right]~, \label{eq:square} \\
	&=& - \beta_\mathrm{eff} \nabla \cdot \hat{ {\bf p}} + \beta_\mathrm{eff} \hat{\mathbb{W}}~.
	\end{eqnarray}
	Here, $\beta_\mathrm{eff} \equiv (D_t + v l_r/2)^{-1}$, $\hat{ {\bf p}} \equiv \rho \nabla V - l_r \boldsymbol{\Gamma}^{(1)}$, and $\hat{\mathbb{W}}$ is the terms in the square brackets of Eq.~\eqref{eq:square}. Using the Green's function of the Laplace operator $G({\bf r}-{\bf r}')$, we write
	\begin{eqnarray}
	\label{Eq:multi} \rho({\bf r}) = \int \mathrm{d} {\bf r}' ~ \beta_\mathrm{eff} G({\bf r} - {\bf r}')  \left[ - \nabla \cdot \hat{ {\bf p}} ( {\bf r}') +  \hat{\mathbb{W}} ( {\bf r}') \right]~.
	\end{eqnarray}
	Since the potential $V({\bf r})$ and the torque $\Gamma({\bf r}, \theta)$ are localized quantities, we can perform multipole expansion of the Green's function in the far-field regime. The multipole order of a term increases with the number of spatial gradients applied on it, and the leading order term in Eq.~\eqref{Eq:multi} has a dipole form corresponding to terms with one spatial derivative. Using the two-dimensional expression for the Green's function, $G({\bf r}) = \ln|{\bf r}|/(2\pi)$, then gives, 
	\begin{equation}
	\rho({\bf r}) = \rho_0 + \frac{\beta_{\rm eff}}{2 \pi} \frac{{\bf r} \cdot {\bf p}}{r^2} + {\cal O}(r^{-2})~,
	\end{equation}
	with
	\begin{equation} \label{Eq:p_torque}
	{\bf p} = - \int \mathrm{d}^2 {\bf r}' \left[ \rho \nabla' V  - {l_r} \boldsymbol{\Gamma}^{(1)} \right]~.
	\end{equation}
	In sum, we find that a localized torque, even in the absence of an
	asymmetric potential, leads to far field of the density and current
	which are identical to those of an asymmetric potential. The only
	difference is a renormalization of the dipole strength.
	
	It is interesting to note that the amplitude of the dipole is
        directly related to the average rate of injection of momentum
        into the system due to both the external potential and the
        rectification of the random active motion by the torques. This
        is very similar to the role played by wall torques in
        destroying the equation of state for the pressure of active
        systems~\cite{solon_pressure_2015,fily2017mechanical}.
	
	We also remark that the results above can be derived straightforwardly
	for a potential $V({\bf r}, \theta)$ which is a function of both ${\bf
		r}$ and $\theta$. In this case, none of the contributions to the
	dipole strength are directly related to the force exerted by the
	potential on the active particles.
	
	\begin{figure} [t]
		\center
		\includegraphics[width=0.9\linewidth]{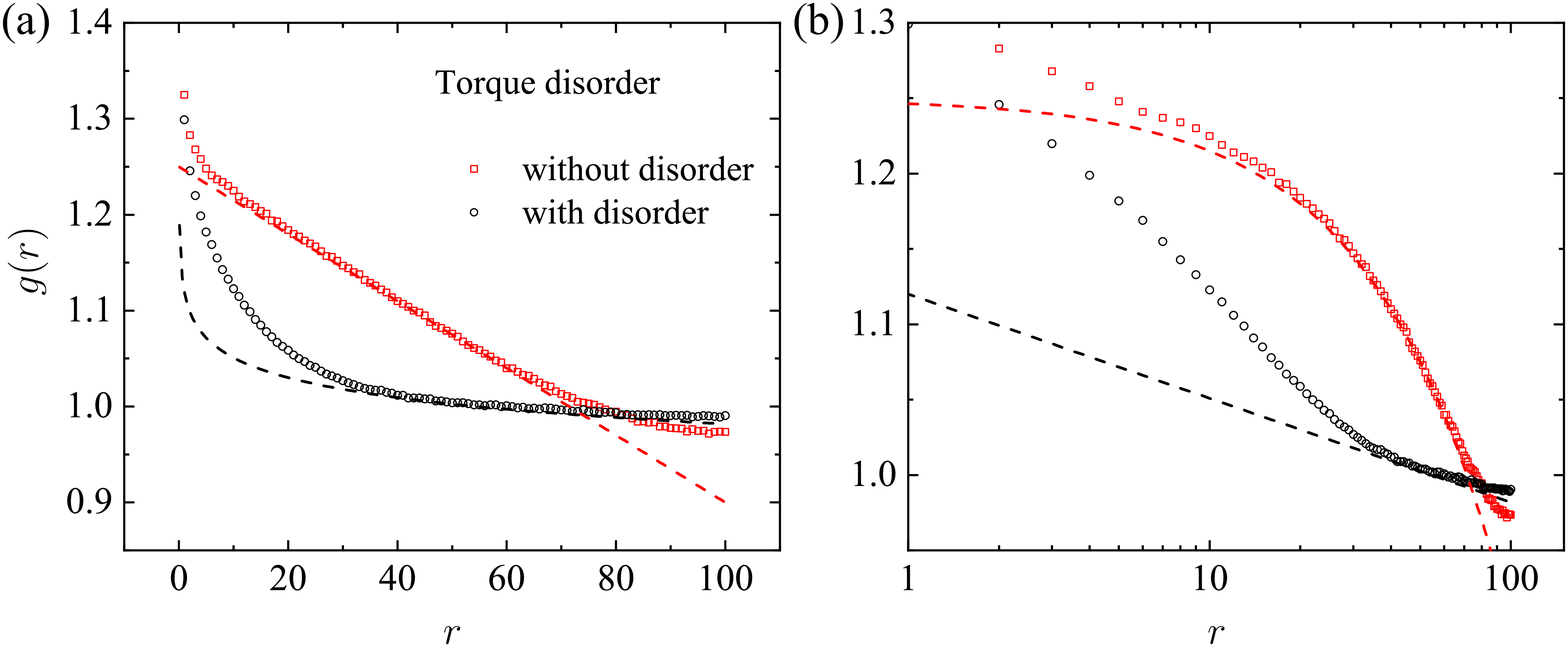}
		\caption{The pair correlation functions shown in (a) a linear scale and (b) a log-linear scale. The simulation results obtained without (red symbols) and with (black symbols) disorder are shown with guidelines presenting linear (red dashed lines) and logarithmic (black dashed lines) decays. Parameters: $L = 200, \Delta V = 3, v = 10, \alpha = 1, \rho_0 = 1$, and $n_M = 2$. 
		}
		\label{Fig3}
	\end{figure}  
	
	To verify that interactions between particles do not change the
	results, we carried out numerical simulations in two dimensions. In
	order to emphasize that even random torques induce the same behavior
	as random potentials, our numerical simulations only consider the
	former and interactions between the particles. The dynamics are
	similar to the rules described in the main text. However, the disorder
	does not affect the translational hopping and instead only modifies
	the tumbling motion.  The 
	details of the
	dynamics are defined as follows. We consider
	run-and-tumble particles with continuous orientation $\hat{e}_\theta =
	(\cos \theta, \sin \theta)$ with $\theta \in [0, 2\pi)$. Due to
	activity, the particles hops from the initial position of a particle
	$\vec{i}$ to one of its neighbors $\vec{j} = \vec{i} + \hat{u}$ with
	rate $W_{\vec{i}, \vec{j}} = \max[v \hat{u} \cdot \hat{e}_\theta,
	0]$ with $v$ controlling the propulsion speed. The steric
	repulsion between the particles modifies the hoping rates as
	$W_{\vec{i}, \vec{j}}^\mathrm{int} = W_{\vec{i}, \vec{j}} (1 -
	n_{\vec j}/n_M)$, where $n_{\vec j}$ is the number of particles at
	$\vec{j}$ and $n_M$ is the maximal particle number per site.  The
	torque disorder biases the particle orientation toward an
	orientations $\theta_{\vec{i}}$, that are uniformly drawn in the
	range $[0, 2 \pi)$ at random for each $\vec{i}$. In doing so, we
	mimic a torque exerted by a potential field of orientation
	$V_{\vec{i}} (\theta)$. To do so a particle at site $\vec{i}$ with
	orientation $\theta$ changes its orientation with the rate $\alpha
	Y_{\vec{i}}(\theta)$, where
	\begin{equation}
	Y_{\vec{i}} (\theta) = \frac{1}{2 \pi} \int \mathrm{d} \theta' ~ e^{V_{\vec{i}}(\theta)- V_{\vec{i}}(\theta')  }~.
	\end{equation}
	The new orientation is chosen with probability density
	${\cal P}(\theta) = e^{-V_{\vec{i}}(\theta)}/(Y_{\vec{i}} (\theta)
	e^{-V_{\vec{i}}(\theta)} )$. In the simulations, we use
	$V_{\vec{i}}(\theta) = - \Delta V \cos^2 (\theta - \theta_{\vec{i}})$.
	
	Figure~\ref{Fig3} shows the results of our numerics. We chose a set of parameters which induce MIPS without disorder, and as a result, the pair correlation function obtained without disorder shown with red symbols in Fig.~\ref{Fig3}(a) presents linear decay, indicating phase separation. On the other hand, the black symbols obtained with torque disorder show logarithmic decay, as verified by the agreement between the black symbols and the black dashed line in Fig.~\ref{Fig3}(b). Thus, our simulation shows that MIPS break-down by the torque disorder, and the resulting disordered phase has the long-ranged density-density correlation similarly to the random potential considered in the main text. 

\section{Description of Supplementary Movies}
We provide movies capturing the time evolution of the particle density
field obtained from simulations. In all the movies, we have $L = 300$,
$v = 13$, $\alpha = 1$, $n_M = 2$, and $\rho_0 = 1$. A detailed
descriptions of each movie is given as follows:
\begin{itemize}
\item \verb@Movie_1_v13_Vx.gif@: These movies show the time evolution of the particle distribution starting from a homogeneous initial condition. The number \verb@x@ specifies the potential strength $\Delta V$.
\item \verb@Movie_2_v13_V7.5_growing.gif@: Here the strength of the disorder, $\Delta V$, is gradually increased from 0 to 7.5. The initial state of the system is phase separated and one can observe how long-range order is lost following the introduction of disorder.
\item \verb@Movie_2_v13_V7.5_vanishing.gif@: In this movie, we gradually reduce the strength of the random potential $\Delta V$ from 7.5 to 0. The initial configuration of the movie is the stationary configuration obtained with $\Delta V = 7.5$. The movie shows how phase separation appears as the disorder is turned off.
\end{itemize}

\bibliographystyle{apsrev4-1}
\bibliography{Disorder_Bib}